# Resolving the Electronic Ground State of La$_3$Ni$_2$O$_{7-\delta}$ Films

Xiaolin Ren[1,2], Ronny Sutarto[3], Xianxin Wu[4], Jianfeng Zhang[1], Hai Huang[5], Tao Xiang[1,2,6], Jiangping Hu[1,2], Riccardo Comin[7,*], X. J. Zhou[1,2,6,8,*], and Zhihai Zhu[1,2,8,*]

[1]*Beijing National Laboratory for Condensed Matter Physics,*

*Institute of Physics, Chinese Academy of Sciences, Beijing 100190, China*

[2]*University of Chinese Academy of Sciences,*

*Beijing 100049, China*

[3]*Canadian Light Source, Saskatoon, Saskatchewan S7N 2V3, Canada*

[4]*CAS Key Laboratory of Theoretical Physics, Institute of Theoretical Physics, Chinese Academy of Sciences, Beijing 100190, China*

[5]*Department of Materials Science, Fudan University, Shanghai, 200433, China*

[6]*Beijing Academy of Quantum Information Sciences, Beijing 100193, China*

[7]*Department of Physics, Massachusetts Institute of Technology, Cambridge, Massachusetts 02139, USA*

[8]*Songshan Lake Materials Laboratory, Dongguan 523808, China*

(Dated: August 05, 2024)

[*]To whom correspondence should be addressed.

Emails: rcomin@mit.edu, XJZhou@iphy.ac.cn, zzh@iphy.ac.cn

**The recent discovery of a superconductivity signature in $La_3Ni_2O_{7-\delta}$ under a pressure of 14 GPa, with a superconducting transition temperature of around 80 K, has attracted considerable attention. An important aspect of investigating electronic structures is discerning the extent to which the electronic ground state of $La_3Ni_2O_{7-\delta}$ resembles the parent state of the cuprate superconductor, a charge transfer insulator with long-range antiferromagnetism. Through X-ray absorption spectroscopy, we have uncovered the crucial influence of oxygen ligands on the electronic ground states of the Ni ions, displaying a charge transfer nature akin to cuprate but with distinct orbital configurations. Both in-plane and out-of-plane Zhang-Rice singlets associated with Ni $d_{x^2-y^2}$ and $d_{z^2}$ orbitals are identified, together with a strong interlayer coupling through inner apical oxygen. Additionally, in $La_3Ni_2O_{7-\delta}$ films, we have detected a superlattice reflection (1/4, 1/4, $L$) at the Ni $L$ absorption edge using resonant X-ray scattering measurements. Further examination of the resonance profile indicates that the reflection originates from the Ni $d$ orbitals. By evaluating the reflection's azimuthal angle dependence, we have confirmed the presence of collinear antiferromagnetic spin ordering and charge-like anisotropy ordered with the same periodicity. Notably, our findings reveal a microscopic relationship between these two components in the temperature dependence of the scattering intensity of the reflection. This investigation enriches our understanding of high-temperature superconductivity in $La_3Ni_2O_{7-\delta}$ under high pressure.**

**INTRODUCTION**

Unconventional superconductors, such as copper-oxide and iron-based compounds, often show superconductivity alongside the destruction of long-range antiferromagnetic ordering when they reach a critical carrier concentration level through doping. As the doping levels increase, the superconducting state may coexist or compete with various electronically ordered phases [1]. For example, cuprate superconductors may exhibit the development of spin and charge order, known as 'stripe' [2-4], while several Fe-based superconductors demonstrate electronic nematic order, breaking discrete rotational symmetry near or within the superconducting phase [5-7]. Understanding the relationship between superconductivity and various emergent electronic orders is crucial for uncovering the mystery of high-temperature superconductivity.

The discovery of the signature of superconductivity in $La_3Ni_2O_{7-\delta}$ under a pressure of 14 GPa with $T_c \approx$ 80 K has garnered considerable interest recently as a promising high-temperature superconductor derived from a transition metal outside copper-oxides or iron-chalcogenides [8]. Furthermore, these bilayer 327-type nickelates, unlike the infinite layer 112-type nickelates, have a completely different electronic configuration compared to cuprates, i.e., they are not engendered by a (half-filled) $d^9$ Mott state but are potentially more complex ground states. However, the research landscape is intricating, with ongoing debates on whether superconductivity is bulk or filamentary [9-12], complicated by the reported intergrowth of different Ruddlesden-Popper (RP) phases [13-16]. Despite the nuances of sample synthesis, substantial theoretical progress has been made in understanding the superconducting mechanism in pressurized $La_3Ni_2O_7$ [17-39]. However, it is essential to note that the orbital configuration of Ni ions and the role of the oxygen ligands in $La_3Ni_2O_7$ remain elusive [40-43]; it is unclear whether it resembles cuprates, which are charge-transfer insulators, or is more akin to infinite-layer nickelates, which are Mott-Hubbard insulators [44-49]. Besides, the potential for electronic ordering, like long-range antiferromagnetic order, has not been fully resolved [50-57].

In the present work, we used X-ray absorption spectroscopy (XAS) at the O $K$ absorption edge and resonant X-ray scattering (RXS) measurements at the Ni $L_3$ absorption edge to investigate the electronic ground state of single-crystalline $La_3Ni_2O_{7-\delta}$ films. In the electronic structure, we have identified in-plane and out-of-plane mobile carrier peaks linked to the Ni $d_{x^2-y^2}$ and $d_{z^2}$ orbitals, along with substantial interlayer coupling facilitated by the inner apical oxygen. Moreover, we have established the presence of collinear antiferromagnetic spin ordering and anisotropic charge-like ordering with a matching periodicity. Notably, the two orders are interconnected yet display distinct temperature-dependent behaviors. These findings offer pivotal insights into the electronic structures of $La_3Ni_2O_7$ and aid in comprehending the underlying mechanism of electron pairings for superconductivity under pressurized conditions.

**RESULTS AND DISCUSSION**

Figure 1a plots a schematic of the crystal structure of $La_3Ni_2O_7$. In Fig. 1b, we present the temperature-dependent resistivity measurements on two representative films of $La_3Ni_2O_{7-\delta}$. These films were synthesized under different oxygen pressures, leading to oxygen variation. The resistivity versus temperature for Film 1 demonstrates insulating behavior. However, for Film 2, the resistivity decreases with decreasing temperature and exhibits an upturn around 80 K, indicating a transition to an insulator. This metal-to-insulator transition is likely due to disorder, given the negligible magnetoresistance across the transition temperature (see supplementary materials for more details). Notably, the resistivity versus temperature for Film 2 resembles that of the bulk crystal, known for its superconductivity under pressure [8]. Since Film 1 was grown with a lower oxygen pressure than Film 2, we anticipate Film 1 to be less stoichiometric than Film 2 regarding oxygen content [40,41,43,58-62]. The stark contrast in resistivity as a function of temperature between the two films may arise from the variation in oxygen content, which could be crucial in understanding the electronic structure and origin of the superconductivity of this material under pressure [8-11,61].

Figures 1c and 1d illustrate the scattering geometry of the RXS experiments. We use a pseudo-cubic notation in defining the unit cell and use reciprocal lattice units (r. l. u.) for the reciprocal space ($H$, $K$, $L$) components, which are given in units of ($2\pi/a$, $2\pi/b$, $2\pi/c$). At $T = 20$ K, the in-plane lattice constants are $a \approx b = 3.789$ Å, and the out-of-plane lattice constant is $c = 20.313$ Å. We have observed superlattice reflections at (1/4, 1/4, 1.9) with photon energy tuned near the Ni $L_3$ absorption edge. This reflection exhibits strong photon polarization dependence, as shown in Fig.1e. It appears much stronger in π- than in σ- polarized incidence geometry, consistent with the recent report on bulk crystal $La_3Ni_2O_{7-\delta}$ [54]. Such a photon polarization dependence suggests that the resonant reflection might be magnetic [54,63].

In Fig. 2a, we present the X-ray absorption spectra (XAS) for both films at the O $K$-edge and the corresponding X-ray linear dichroism (XLD) to uncover the orbital configurations of the Ni ions and the influence of oxygen ligands. According to the Zaanen-Sawatzky-Allen (ZSA) scheme, which classifies insulators into Mott-Hubbard and charge-transfer types, the role of ligands in determining the smallest energy gap of the charge excitations becomes evident [62]. The charge-transfer nature manifests as a pre-edge peak in the oxygen $K$-edge XAS [64]. We observe clear pre-edge peaks in O $K$-edge XAS in both films around a photon energy of ~528 eV. They exhibit apparent polarization dependence: a sharp peak with in-plane polarization and two broad peaks with out-of-plane polarization. This indicates that the oxygen ligands significantly influence the electronic ground state of Ni cations. Notably, the XLD analysis of the difference of pre-edge peaks with different polarization enables us to deduce the Ni 3$d$ orbital configurations through $p$-$d$ hybridization within the $NiO_2$ bilayers. As shown in Fig. 2a, the main peak at

~528 eV with the in-plane polarization shows little sample dependence, consistent with $d_{x^2-y^2} - p_{x,y}$ Zhang-Rice singlet state in cuprate. Intriguingly, the two dips flanking the main peak with the out-of-plane polarization exhibit slight sample dependence, reflecting the strong hybridization of the inner apical O $p_z$ and the Ni $d_{3z^2-r^2}$ orbital and the formation of an out-of-plane $d_{3z^2-r^2}$ Zhang-Rice singlet state in bilayer nickelates. The energy difference of ~1.2 eV between the two dips is approximately two times that for the interlayer hopping of the $d_{3z^2-r^2}$ orbitals revealed from the density functional calculations [8,65,66]. Thus, these two dips correspond to the $d_{3z^2-r^2}$ interlayer bonding and antibonding states, as illustrated in Fig. 2b. This contrasts with the recent angle-resolved photoemission spectroscopy (ARPES) measurements on a bulk crystal La$_3$Ni$_2$O$_{7-\delta}$, where the correlation effects were proposed to push the bonding states derived from the $d_{3z^2-r^2}$ orbitals downward by about 50 meV below the Fermi level [66]. Such discrepancy may arise from the variance in oxygen concentration and its impact on covalency (mainly through the apical oxygen) and/or the Fermi level position and the external pressure from substrates. As shown in Fig. 2a, the energy of the bonding states from the $d_{3z^2-r^2}$ orbitals appears to be more sensitive to the possible oxygen variation compared with the antibonding state from the $d_{3z^2-r^2}$ orbitals and the bands from the $d_{x^2-y^2}$[67], consistent with the observed prominent inner apical oxygen vacancy [43].

We conducted momentum scans along high symmetry directions in momentum space for a comprehensive search of translational symmetry-breaking signatures. This led us to the observation of a superlattice peak (SP) at the wave vector $Q_{SP}$ = (1/4, 1/4, 1.9) in our La$_3$Ni$_2$O$_{7-\delta}$ films, as shown in Fig. 2c. To further understand the nature of this reflection, we performed momentum scans across the $Q_{SP}$ with various photon energies. The results, depicted in Fig. 2c, demonstrate the resonant enhancement of the scattering signals, with a more pronounced increase in the scattered signal when using π-polarized incident photons, suggesting the magnetic nature of the reflection. In Fig. 2d, we present the energy scans with wave vectors fixed at $Q_{SP}$, which reproduce the energy profiles in Fig. 2c. To provide a reference, we also repeated the energy scans slightly off this wave vector to characterize the scattering background under an identical geometry. The spectra taken slightly off the $Q_{SP}$ exhibit much weaker signals (primarily resulting from fluorescence) than those measured at the wave vector, confirming the electronic nature of the reflection at $Q_{SP}$.

Figures 3a and 3b summarize the temperature dependence of the scattered intensity of the reflection at $Q_{SP}$ with π- and σ-polarized incident photons, respectively. Fig. 3a reveals a clear transition at $T \approx 150$ K for both films. Figures 3c-d show the representative scans using π-polarized incident photons at various temperatures, confirming vanishing intensity above 150 K. This transition at $T = 150$ K aligns with the earlier studies [42,52-55], suggesting the transition from magnetic ordering. However, surprisingly, the measurements with σ-polarized incident photons show different characteristics. The temperature dependence of the scattered signals takes an entirely different turn, exhibiting three characteristic temperatures near $T \approx 200$ K, 150 K, and 110 K, respectively. The corresponding representative scans in Fig. 3e-f show the vanishing intensity near $T \approx 200$ K. This unexpected behavior is a fascinating twist in our understanding. While the transition at $T = 150$ K likely mirrors the same transition as revealed in the π channel, the transition at $T = 200$ K suggests a different type of order with the same wave vector that already occurs about 50 K above the proposed magnetic order at $T = 150$ K. Furthermore, the peak intensity starts decreasing at $T \approx 110$ K as decreasing temperature, in sharp contrast to that measured in the π channel.

The discrepancy in the measurements of the scattering intensity versus temperature using the π- and σ-polarized incident photons indicates that the superlattice reflection at $Q_{SP}$ may contain multiple components. Therefore, we conducted a symmetry-restricted tensorial examination of the azimuthal angle dependence of the scattered intensity to resolve the electronic texture underlying the superlattice reflection. The scattered intensity of the resonant reflection is given below [68-70],

$$I = \left|\sum_j e^{i(\vec{k}_{in}-\vec{k}_{out})\cdot\vec{r}_j} \varepsilon_{out}^* \cdot F_j \cdot \varepsilon_{in}\right|^2, \qquad (1)$$

where $\varepsilon_{in(out)}$ represents the polarization of incoming (outgoing) photons, and $\vec{k}_{in(out)}$ stands for the wave vector of the incoming (outgoing) photons. $\vec{r}_j$ and $F_j$ stand for the position of the atom $j$ and its scattering tensor, respectively. The polarization of the incident photons is resolved between the σ- and π- channel in the measurements, while outgoing photons are not polarization resolved. Therefore, we measured $I_{\pi\sigma} + I_{\pi\pi}$ when using π-polarized incoming photons, and $I_{\sigma\pi} + I_{\sigma\sigma}$ when using σ-polarized incoming photons.

We propose different ordered electronic motifs with an in-plane wavevector (1/4, 1/4) aligned with the wavevector revealed in the measurements. In Fig. 4a, we enforce the antiparallel alignment of next-nearest-neighboring spins while allowing the nearest-neighboring spins to align randomly except for being antiparallel; this comprises both collinear (double spin stripe) and non-collinear antiferromagnetism. In Fig. 4b, we propose a spin-charge stripe pattern that enforces the antiparallel alignment of next-nearest-neighboring spins with spinless charge domains; this would imply charge order with an in-plane wavevector (1/2, 1/2), double that of spin order. Figure 4c shows the measurements for the scattering intensity ratio π/σ as a function of the azimuthal angle. It turns out that the model of the spin-charge stripe cannot give a satisfactory fit to the measurements. The dashed curve represents the modeling of the spin-charge stripe with the magnetic moment pointing along the (1 -1 0) direction, as illustrated in Fig. 4b. This model substantially deviates from the measurements. We then explore a model with the spin textures as illustrated in Fig. 4a, and find both the non-collinear magnetic structures and double spin stripe yield a better agreement with the data, as indicated by the solid red curve from the fit to the data (see the Supplementary materials for further information). However, it still significantly deviates from the observed intensity ratios. Considering that the models of both collinear and complex magnetic spin textures cannot give a satisfactory fit to the data, we expect the likely presence of an anisotropic charge distribution that further lowers the symmetry on the spinless sublattices, as illustrated with green ellipses in Fig. 4b. A fit using the model that consists of both charge anisotropy and collinear antiferromagnetic spin order is given by the solid light blue curve in Fig. 4c, exhibiting a significantly improved agreement to the measurement, yielding the collinear antiferromagnetic spin ordering with spin moment primarily pointing along the (1 -1 0) orientation (see the Supplementary Materials for details).

The discovery of two components contributing to the resonant reflection at (1/4, 1/4) from the symmetry-restricted tensorial analysis naturally explains the remarkable differences in the scattering intensity measurements versus temperature when using σ- and π-polarized incident photons. In particular, the scattering tensor for the charge anisotropy contains only the diagonal components dominating the σ-σ channel. The scattering intensity versus temperature measured using the σ-polarized incoming light mainly reflects the presence of charge anisotropy that emerges around $T \approx 200$ K, followed by the collinear spin ordering at $T \approx 150$ K. Notably, the transition at $T \approx 150$ K agrees with earlier studies using various experimental techniques [42,52-55,57]. The sudden decrease in scattering intensity at $T \approx 110$ K coincides with the characteristic temperature of the anomaly observed in various measurements [42,51], which is proposed to originate from a density wave intimately linked to the orbital-derived Fermi surface topology [51].

In summary, we have revealed the electronic motif and the corresponding thermal dynamics in oxygen-varying films of $La_3Ni_2O_{7-\delta}$ using X-ray absorption spectroscopy and resonant X-ray scattering. We have uncovered the pivotal role of oxygen ligands, which hybridize with the Ni 3*d* orbitals, forming the Zhang-Rice-like bands. However, unlike cuprates in which only the hybridization between O-2$p$ and Cu $3d_{x^2-y^2}$ dominates the lower energy physics, in our La$_3$Ni$_2$O$_{7-\delta}$ films, both the Ni $3d_{x^2-y^2}$ and $3d_{3z^2-r^2}$ hybridize with the in-plane O-2$p_{x,y}$ orbitals, and apical O-2$p_z$ orbitals, as manifested by the pre-edge peak at the O $K$ edge XAS, suggesting both the Ni-$3d_{x^2-y^2}$ and $3d_{3z^2-r^2}$ play essential roles in the low-energy physics in La$_3$Ni$_2$O$_{7-\delta}$ [8,17,19-21,31].

The revelation of the possible spin ordering at (1/4, 1/4) at $T \approx 150$ K in our film samples aligns with the recent studies on the bulk crystals [54]. More subtly, our azimuthal angle measurements, along with the symmetry-restricted tensorial analysis, suggest the presence of an aspherical charge distribution at the sublattices of the spinless sites in the spin-charge model resulting in two components in the reflection at (1/4, 1/4). The ordering due to the charge anisotropy exhibits two characteristic temperatures at $T \approx 200$ K and $T \approx 110$ K. Note that our symmetry-restricted tensorial analysis cannot pinpoint the source accounting for the transition at $T \approx 200$ K since any ordering of charge-like anisotropy would assume a similar format of a diagonal matric tensor. One possibility is that an ordered Jahn-Teller effect may exist on the spinless sublattices [71], leading to charge anisotropy. Another possibility is the presence of a magnetic quadrupole ordering on the spinless sites, which is compatible with our symmetry analysis (see the supplementary materials for further information).

Studying the electronic ground state of La$_3$Ni$_2$O$_{7-\delta}$ films with varying oxygen concentrations could have important implications for understanding the superconductivity discovered under pressurized conditions. In unconventional superconductors, various orderings, such as spin order, charge order, and spin-charge stripe, often appear to compete with superconductivity. Perturbations such as applying strain, pressure, or magnetic field can disrupt the balance between the superconducting phase and other competing orders. This scenario may also apply to La$_3$Ni$_2$O$_{7-\delta}$, where high pressure could suppress the spin-charge stripe while enhancing superconductivity. Notably, the resonant reflection at $Q_{SP}$ in the film grown at lower oxygen pressure exhibits a much higher intensity than that in the film grown at higher oxygen pressure. The resistivity versus temperature in the high oxygenated film resembles that of the bulk samples which become superconducting under pressure. Variations in oxygen levels may have already suppressed the spin-charge ordering to some extent, thereby aiding pressure in eliminating it and enabling the full emergence of superconductivity.

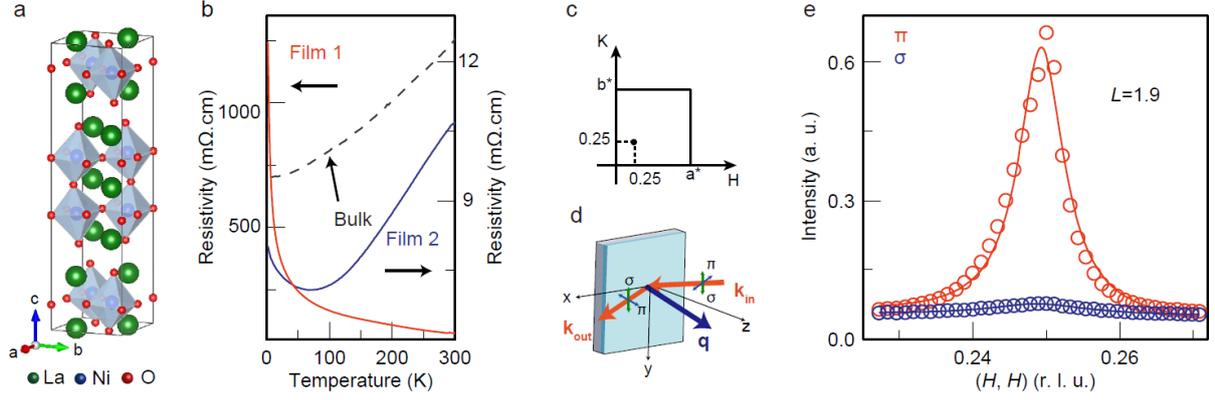

**FIG. 1: Superlattice reflection with wave vector at (1/4, 1/4, *L*) in La$_3$Ni$_2$O$_7$. a**, Crystal structure of La$_3$Ni$_2$O$_7$ with lattice parameters $a \approx b = 3.789$ Å and $c = 20.313$ Å. The green, blue, and red solid circles represent La, Ni, and O atoms, respectively. **b**, The temperature-dependent resistivity for two typical films of La$_3$Ni$_2$O$_{7-\delta}$; the dashed line represents the resistivity measurement of a La$_3$Ni$_2$O$_7$ single crystal in ref. [8] and is rescaled for clarity. **c**, Reciprocal space map of the position of the resonant diffraction peaks in La$_3$Ni$_2$O$_7$. **d**, A schematic of the RXS scattering geometry. **e**, Representative momentum scans across the wave vector (1/4, 1/4, 1.9) measured at $T = 20$ K and $E = 852.2$ eV with π polarized (red open circle) and σ polarized (blue open circle) incident photons. The solid lines fit the data using the Lorentz profiles and a linear background.

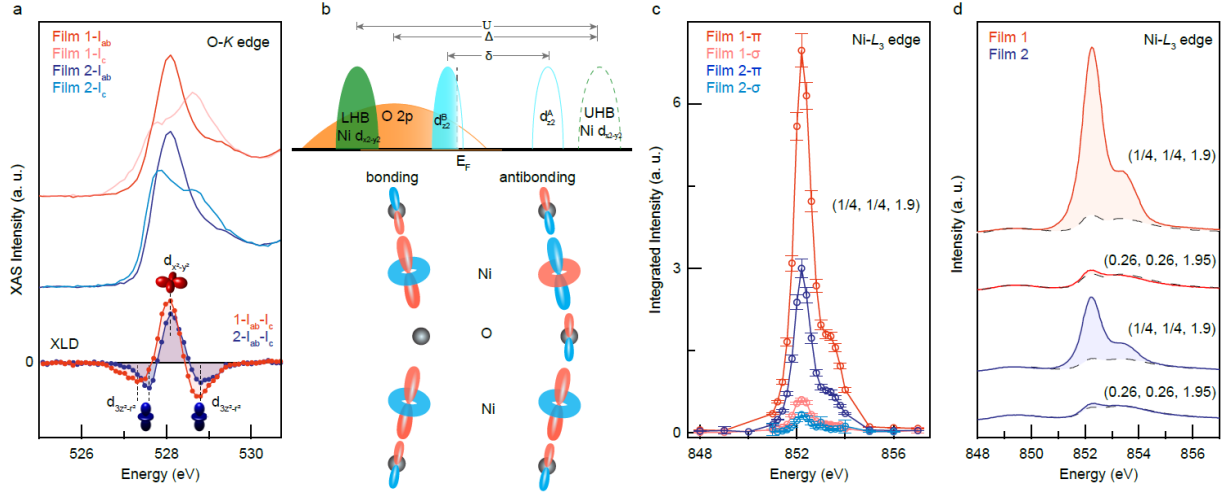

**FIG. 2: X-ray absorption spectra (XAS), and energy profiles of the resonant reflections in $La_3Ni_2O_{7-\delta}$. a,** The X-ray absorption spectra (XAS) and X-ray linear dichroism (XLD) near the O-$K$ edge for Film 1 (red color) and Film 2 (blue color). The XAS are offset for clarity. **b,** Schematic density of states of $La_3Ni_2O_7$ and the corresponding bonding and antibonding orbitals. **c,** Energy profiles of the resonant reflection at (1/4, 1/4, 1.9) near the Ni-$L_3$ absorption edge with π- and σ- polarizations, and the peak intensity is maximized at $E = 852.2$ eV. **d,** Energy scans at constant wavevector on and off the $Q_{SP}$ = (1/4, 1/4, 1.9) using different photon polarizations for two $La_3Ni_2O_{7-\delta}$ films. The solid lines corresponds to the energy scans on $Q_{SP}$, while the dashed lines correspond to that off the $Q_{SP}$. The curves are offset for clarity.

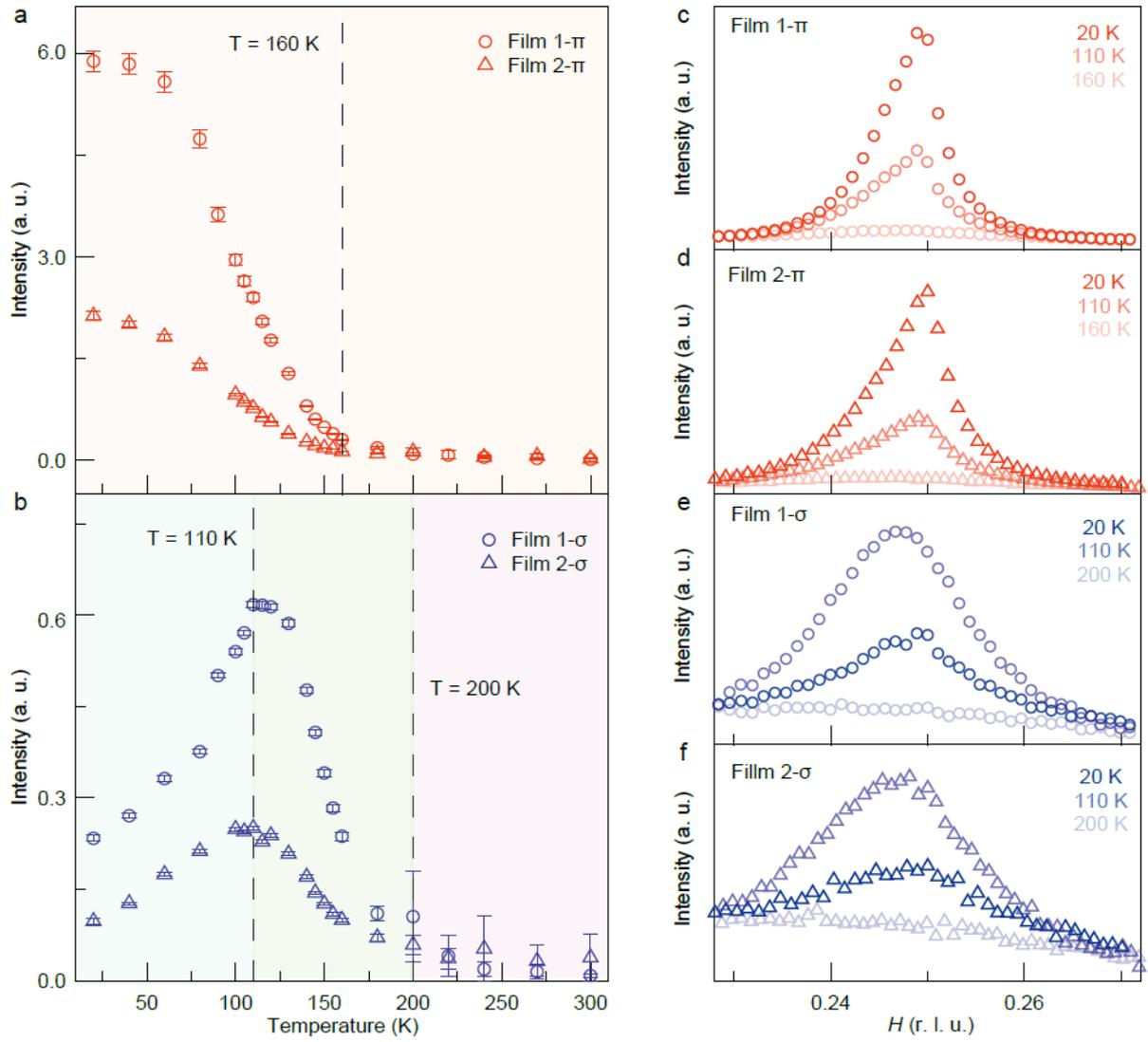

**FIG. 3: Temperature dependence of the intensity of the resonant peak at (1/4, 1/4, 1.9).** **a**, **b**, The diffraction peak intensity as a function of temperature taken with π (**a**) and σ (**b**) polarized incident light at 852.2 eV for the $La_3Ni_2O_{7-\delta}$ films. The dashed lines mark the transition temperature. **c, d (e, f)**, Representative momentum scans across the wave vector (1/4, 1/4, 1.9) for Film 1 and Film 2 measured with π (σ) polarization at various temperatures. The circle and triangle correspond to Film 1 and Film 2, respectively.

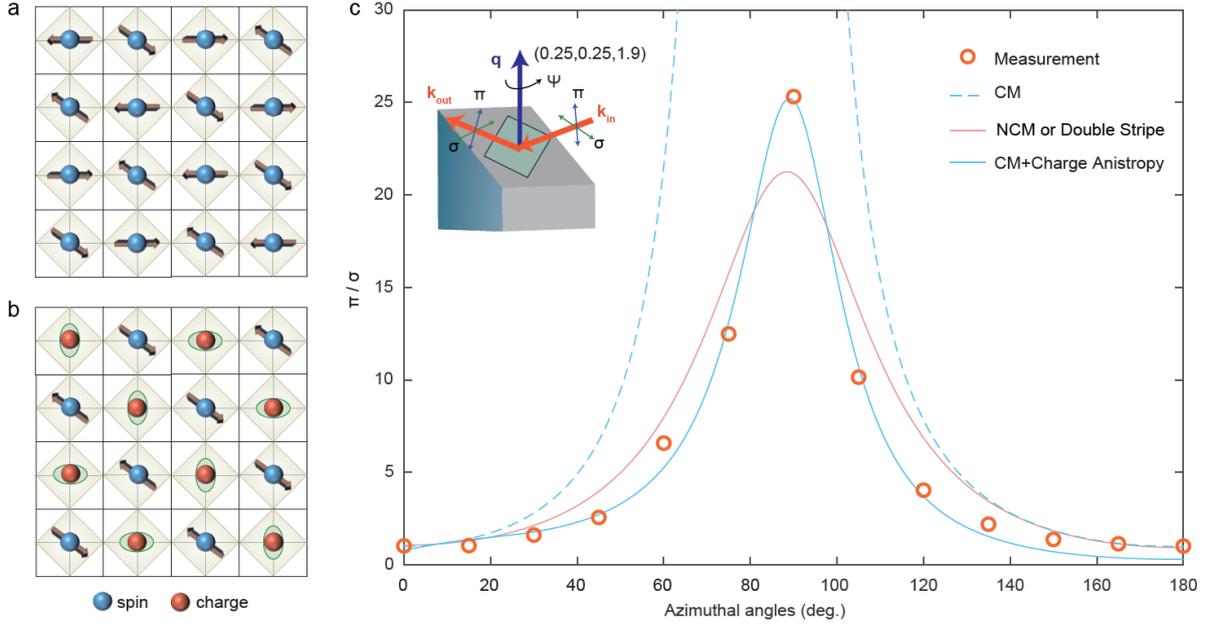

FIG. 4: Azimuthal angle dependence of the resonant reflection at (1/4, 1/4, 1.9) in La$_3$Ni$_2$O$_{7-\delta}$. **a**, **b**, A schematic of spin configuration illustrating the magnetic structure in real space. **a** shows spin configuration that can be both non-collinear antiferromagnetism (NCM) and collinear antiferromagnetism (CM). **b** shows spin-charge stripe configurations with charge anisotropy on the spinless sublattices (CM + charge anisotropy). The blue and red solid circles represent the spin and charge sublattices, respectively; brown arrows denote the spin momentum directions. The green shaded areas indicate the presence of charge anisotropy lowering the local symmetry on the spinless sublattices. **c**, Azimuthal dependence of the integrated intensity of the reflection (1/4, 1/4, 1.9) at 20 K (where not shown, the error bars are within the symbol size). The azimuthal angle $\psi = 0$ corresponds to a sample orientation where the (1 1 0) orientation lies within the scattering plane. The solid blue lines represent the best fit for a model illustrated in (**b**), while the solid red line corresponds to the best fit illustrated in (**a**) corresponding to noncollinear spin configurations. The model of the spin-charge stripe without charge anisotropy does not seem to converge in the fit regardless of the initial parameters; the dashed line represents the simulation of the spin-charge stripe with the spin momentum direction fixed to the (1 -1 0)

orientation. The inset in **c** shows a schematic of the scattering geometry of azimuthal angle-dependent measurements.

## Methods

**Sample preparation.** Thin films of the precursor phase $La_3Ni_2O_7$ with a thickness of ~30 nm were prepared using pulsed laser deposition (PLD) on (001)-oriented $LaAlO_3$ substrates with a 248-nm KrF excimer laser. The substrate was kept at 750 °C during growth under an oxygen partial pressure of 150 mTorr. After deposition, the films were cooled to room temperature at 5 °C per minute in the oxygen partial pressure of 100 Torr.

**XAS and RXS measurements.** The X-ray absorption spectroscopy (XAS) and resonant x-ray scattering (RXS) measurements at the Ni *L*-edge and O-*K* edge were performed at the REIXS (10ID-2) beamline of the Canadian Light Source (CLS). At these photon energy ranges, the beamline provides a photon flux of about $5\times10^{11}$ photons per second with energy resolution $\Delta E/E \sim 2\times10^{-4}$. An in-vacuum 4-circle diffractometer hosts the endstation under an ultra-high vacuum (UHV) condition below $5\times10^{-10}$ mbar. The XAS spectra were collected in both total fluorescence yield (TFY) and total electron yield (TEY) modes at 20 K with linear vertical ($\sigma$) and horizontal ($\pi$) light polarizations. The XAS spectra are normalized to the incident photon flux. The RXS spectra are extracted after a Lorentzian fitting with a linear background.

## Acknowledgment

This work was supported in part by the National Key Research and Development Program of China (Grant No. 2022YFA1403900 and 2021YFA1401800), the National Natural Science Foundation of China (Grant No. 12074411), the Strategic Priority Research Program (B) of the Chinese Academy of Sciences (Grant No. XDB25000000), and the Synergetic Extreme Condition User Facility (SECUF). Part of the research described in this paper was performed at the Canadian Light Source, a national research facility of the University of Saskatchewan, which is supported by the Canada Foundation for Innovation (CFI), the Natural Sciences and Engineering Research Council (NSERC), the Canadian Institutes of Health Research (CIHR), the Government of Saskatchewan, and the University of Saskatchewan.

**Author contributions:** Z.H.Z. and X.J.Z. conceived the research. X.L.R. prepared and characterized the film samples. X.L.R., R.S., and Z.H.Z. performed the RXS experiments. X.L.R., Z.H.Z, R.C., and H.H. analyzed the experimental data. T.X., J.P.H., X.X.W., and J.F.Z. provided theoretical guidance with DFT calculations. X.L.R. and Z.H.Z. wrote the manuscript with the input of all co-authors.

# Supplementary Material for

# Resolving the electronic ground state of La$_3$Ni$_2$O$_{7-\delta}$ films

Xiaolin Ren[1,2], Ronny Sutarto[3], Xianxin Wu[4], Jianfeng Zhang[1], Hai Huang[5], Tao Xiang[1,2,6], Jiangping Hu[1,2], Riccardo Comin[7,*], X. J. Zhou[1,2,6,8,*], and Zhihai Zhu[1,2,8,*]

[1]*Beijing National Laboratory for Condensed Matter Physics,*
*Institute of Physics, Chinese Academy of Sciences, Beijing 100190, China*

[2]*University of Chinese Academy of Sciences,*
*Beijing 100049, China*

[3]*Canadian Light Source, Saskatoon, Saskatchewan S7N 2V3, Canada*

[4]*CAS Key Laboratory of Theoretical Physics, Institute of Theoretical Physics, Chinese Academy of Sciences, Beijing 100190, China*

[5]*Department of Materials Science, Fudan University, Shanghai, 200433, China*

[6]*Beijing Academy of Quantum Information Sciences, Beijing 100193, China*

[7]*Department of Physics, Massachusetts Institute of Technology, Cambridge, Massachusetts 02139, USA*

[8]*Songshan Lake Materials Laboratory, Dongguan 523808, China*

(Dated: August 05, 2024)

[*]To whom correspondence should be addressed.

Emails: rcomin@mit.edu, XJZhou@iphy.ac.cn, zzh@iphy.ac.cn

## Film Growth and Characterizations

The $La_3Ni_2O_7$ films with a thickness of about 30 nm were grown by the PLD using a 248-nm KrF excimer laser (COMPex 201, Coherent). During growth, the $LaAlO_3$ substrate temperature was maintained at 750 °C, and the oxygen partial pressure was 150 mTorr for Film 1 and 200 mTorr for Film 2, respectively. After deposition, the films were cooled to room temperature at a rate of 5 °C/min with an oxygen partial pressure of 100 Torr. The pulse-to-pulse repetition rate was set to 4 Hz, and the pulse laser energy was set to 500 mJ. The laser target is a chemically measured mixture containing $La_2O_3$ (Alfa Aesar, 99.99%) and NiO (Sigma Aldrict-Chemie GmbH, 99.995%) prepared by a solid-state reaction, the pellets were heated in air at three °C/min to 1100 °C and held for 12 h.

Figure S1a shows a 2θ-ω scan of the out-of-plane XRD patterns for two typical $La_3Ni_2O_{7-\delta}$ films grown on $LaAlO_3$ (001) substrates, identifying the right structural phase, i.e., Ruddlesden–Popper (RP) perovskites of $La_3Ni_2O_{7-\delta}$. Film 1 has a smaller *c* lattice constant (20.3126 Å) than Film 2 (21.1105 Å).   Figs. S1b and 1c show the reciprocal space map (RSM) around the (1 0 17) peaks of Film 1 and Film 2, respectively, confirming that the films are epitaxially grown on $LaAlO_3$ substrates and coherently strained by the substrates. Fig. S2 shows the resistivity versus temperature under various magnetic fields for Film 2. A metal-to-insulator transition occurs around 80 K, but it does not show any response to the applied magnetic field. It suggests the meal-to-insulator transition likely arises from the disorder.

## STEM Characterizations

In the scanning transmission electron microscopy (STEM) images in Fig. S3, the distinction between atoms of different brightness, based on the difference in atomic number, is evident. The interface between the substrate and the two films is clearly visible. By comparing the high-angle annular dark field (HAADF) STEM images shown in Fig. S3a and 3c and Annular Bright Field (ABF) STEM images shown in Fig. S3b and S3d, we find that the perovskite structure of the substrate and the RP structure of the film are orderly arranged along the *c*-axis without any noticeable defects or other phases. Despite the different growth oxygen partial pressure along with the difference in transport measurements on the two films, no discernible difference in microstructure is observed. However, we seem to have a poorer resolution, which does not allow us to detect the likely presence of oxygen defects as reported in bulk samples [1].

**X-ray absorption spectra collected at the La *M*-edge and Ni *L*-edge on La$_3$Ni$_2$O$_{7-\delta}$ films**

Figure S4 shows the normalized X-ray absorption spectra (XAS) collected via the total electron yield (TEY) mode at the La *M*-edge and the Ni *L*-edge for Film 1 and Film 2, the same films studied in the main text. The salient peaks at ~833 eV and ~850 eV correspond to the La $M_5$ ($3d_{5/2}\rightarrow 4f$) and the La $M_4$ ($3d_{3/2}\rightarrow 4f$) absorption edges, respectively, while the relatively weaker peaks at ~852 eV and ~870 eV correspond to the Ni $L_3$ ($2p_{3/2}\rightarrow 3d$) and the Ni $L_2$ ($2p_{1/2}\rightarrow 3d$) absorption edges, respectively. The energy difference between the peak maximum of the XAS at the La $M_4$ and the Ni $L_3$ is about 2 eV. The superlattice peak at the wave vector (1/4, 1/4, 1.9) in La$_3$Ni$_2$O$_{7-\delta}$ appears only at the Ni $L_3$ absorption edge, in sharp contrast to the infinite-layer nickelates in which superlattice peaks show up at both the Ni *L* edges and the rare earth (La, Nd, and Pr) *M* edges [2-6]. The double-peak feature at the Ni $L_3$ absorption edge resembles that measured from the bulk crystal [7]. In analogous to the cuprates, the double peaks may be ascribed to different final states of the Ni ions: Ni-$2p^53d^8$ and Ni-$2p^53d^8L$, where *L* denotes the oxygen ligands. By performing X-ray linear dichroism (XLD) analysis, we find that the out-of-plane XAS peak has a higher intensity than that of in-plane XAS at ~852.2 eV, which indicates that the holes primarily occupy the $3d_{3z^2-r^2}$ orbital. In contrast, the out-of-plane XAS peak shows a lower intensity than the in-plane XAS peak at ~853.4 eV, suggesting the holes primarily occupy the $3d_{x^2-y^2}$ orbital when the oxygen ligands are at play. Overall, the XAS for two La$_3$Ni$_2$O$_{7-\delta}$ films appears similar, implying a nearly identical electronic configuration of the Ni 3$d$ orbital configurations for the two films.

### Density of States of bulk La$_3$Ni$_2$O$_7$

Fig.S5 shows the orbital-resolved density of states (DOS) for bulk La$_3$Ni$_2$O$_7$. It is apparent that the low-energy physics is dominated by the $d_{x^2-y^2}$ and $d_{z^2}$ orbitals and they exhibit strong hybridization with O $p_{x,y}$ and $p_z$ orbitals. The strong interlayer coupling between $d_{z^2}$ orbital through the inner apical oxygen generates bonding and antibonding states. The bonding state is almost fully filled, while the antibonding state is empty. These orbital characteristics and hybridization can account for the pre-edge XAS peaks at the O-$K$ edge shown in Fig. 2a.

### $L$ dependence of the charge order in La$_3$Ni$_2$O$_7$ films

In Fig. S6, we show the $L$ dependence of the scattering intensity across the wavevector (1/4, 1/4, $L$) in the La$_3$Ni$_2$O$_{7-\delta}$ films. The measurements were taken at $T$ = 20 K with the incident photons being $\pi$ polarized and the energy $E$ = 852.2 eV. Fig. S6b represents the momentum scans across the wave vector (1/4, 1/4, $L$) for Film 1 (red) and Film 2 (blue). For both films, the superlattice peak exhibits weak but definite $L$ dependence with a maximum at $L$～1.9. In Fig. S6a, we also show the momentum scan along $L$ for Film 1, exhibiting a similar weak $L$ dependence of the peak intensity. By fitting the data with the Gaussian profile (black solid line), we obtain the correlation length $\xi \approx$ 2.89 Å along $L$, suggesting a quasi-2D nature of the reflection at (1/4, 1/4, $L$～1.9).

### Symmetry-restricted tensorial analysis.

As illustrated in Fig. 4a and 4b in the main text, we can define structure factors as $f_{i,j}$, where $(i, j)$ labels the atomic sites on the square lattice. Here, we do not consider the ordering pattern normal to the *a-b* plane since the reflections are observed at the wave vector $\mathbf{Q} = (1/4, 1/4, L{\sim}2)$. The structure factor at $\mathbf{Q}$ can be written as $F = \sum_{i,j} e^{i\vec{k}\cdot\vec{r}_{i,j}} f_{i,j}$. In the model, as illustrated in Fig. 4a, we assume $f_{i,j} = -f_{i\pm 2,j}$, $f_{i,j} = -f_{i,j\pm 2}$, and $f_{i,j} = f_{j,i}$. These restrictions cover three typical spin patterns: spin-charge stripe, double spin stripe, and non-collinear spin ordering. We can label the structure factors $F_1$, $F_2$, $F_3$, and $F_4$, respectively, at the atomic positions $(i,j) = (0,0), (1,0), (2,0)$, and $(3,0)$. The total structure factor F is calculated as $F = 4(F_1 + iF_2 - F_3 - iF_4)$, which can be simplified to $F = 8(F_1 + iF_2)$ with $F_1 = -F_3, F_2 = -F_4$. The atomic scattering tensor can be written as

$$F_i = i d_i^{(1)} \begin{pmatrix} 0 & m_z^i & -m_y^i \\ -m_z^i & 0 & m_x^i \\ m_y^i & -m_x^i & 0 \end{pmatrix}, \qquad (1)$$

where $d_i^1$ represents the coefficient for the pure magnetic term and depends on the photon energy. In Fig. 4b, we introduce charge anisotropy in the spinless sublattices. The scattering tensor for the anisotropic charge distribution can be written as

$$F_i = d_i^{(2)} \begin{pmatrix} a_i & 0 & 0 \\ 0 & b_i & 0 \\ 0 & 0 & c_i \end{pmatrix}, \qquad (2)$$

where coefficient $d_i^2$ is a function of photon energy and a real value. This diagonal tensor can represent the anisotropic charge distribution arising from magnetic quadrupolar contribution. The total structural factor $F = 4(F_1 + iF_2 - F_3 - iF_4)$ is written as

$$F = 8id_1^{(1)} \begin{pmatrix} 0 & m_z^1 & -m_y^1 \\ -m_z^1 & 0 & m_x^1 \\ m_y^1 & -m_x^1 & 0 \end{pmatrix} + 4id_2^{(2)} \begin{pmatrix} a_2 - a_4 & 0 & 0 \\ 0 & b_2 - b_4 & 0 \\ 0 & 0 & 0 \end{pmatrix}, \quad (3)$$

where we have assumed that symmetry breaking occurs primarily in the ab-plane by setting $c_2 - c_4 = 0$.

We can define the magnetic momentum $(m_x^i, m_y^i, m_z^i)$ by spherical coordinates, $(m_x^i = m^i \cos(\phi)\sin(\theta), m_y^i = m^i \sin(\phi)\sin(\theta), m_z^i = m^i \cos(\theta)$, where $m^i$ represents the magnitude of the spin moment, and $\theta$ and $\phi$ represents angles of the spin moment relative to the crystallographic axes $c$ and $a$, respectively. The light polarization is defined in the crystallographic coordinate frame involving the scattering vector $\mathbf{Q} = \mathbf{k}_{out} - \mathbf{k}_{in}$ and a second arbitrary vector that must be perpendicular to $\mathbf{Q}$, named as $\mathbf{Q}_\perp$. We have the following formula for the $\mathbf{k}$ vectors and photon polarizations σ and π [8]:

$$k_{in} \parallel \sin(\theta')\left[\cos(\psi)\hat{Q}_\perp + \sin(\theta)(\hat{Q}_\perp \times \hat{Q})\right] + \cos(\theta)\hat{Q},$$

$$k_{out} \parallel \sin(\theta')\left[\cos(\psi)\hat{Q}_\perp + \sin(\theta)(\hat{Q}_\perp \times \hat{Q})\right] - \cos(\theta)\hat{Q},$$

$$\sigma = (\hat{k}_{in} \times \hat{k}_{out})/\sin(2\theta'), \quad \pi_{in} = (\hat{k}_{in} \times \sigma), \quad \pi_{out} = (\hat{k}_{out} \times \sigma), \quad (4)$$

Where $\theta'$ presents the angle between $\mathbf{Q}$ and $\mathbf{k}_{in}$, and $\psi$ is the azimuthal angle. The scattered intensity of the resonant reflection is given:

$$I_{scat} = \left|\sum_j e^{i(\mathbf{Q}\cdot\mathbf{r}_j)} \varepsilon_{out}^* \cdot F_j \cdot \varepsilon_{in}\right|^2, \quad (5)$$

Where $\varepsilon_{out}^*(\varepsilon_{in})$ represents the photon polarizations whose formulas are given in eq. (4).

The spin-charge stripe does not fit the data well, so we explored the complex magnetic structure, double spin stripe, and a combination of spin-charge stripe and charge-like anisotropy. We found that selecting different initial parameters in our fitting process resulted in a good fit in two different regions of the parameter spaces, corresponding to two typical spin textures of noncollinear antiferromagnetism and double spin stripe (shown in Figs. S7a, b). The best-fit results for these two spin textures are shown in Fig. S7 as the solid and dashed lines. The fitting parameters (m, ϕ, θ) for the spin moments are (1, 315.986°, 90.814°) and (0.121, 47.975°, 46.915°) for the solid line in Fig. S7a, and (0.987, 320.23°, 88.582°) and (0.7197, 307.85°, 95.228°) for the dashed line in Fig. S7b. However, these fitting parameters may vary depending on the initial parameters chosen in the nonlinear fit. After numerous attempts, we found that using non-collinear antiferromagnetic and double spin stripe spin models did not improve the fit shown in Fig. S7c. While they fit the data better than the spin-charge stripe model, they still substantially deviate from the measurements. The fit using a combination of spin-charge stripe and charge-like anisotropy significantly improved the agreement with the data, as shown in Fig. 4c. In the best fit to the data set, as shown by the solid blue curve in Fig. 4c in the main text, we found $(a_2 - a_4)/(b_2 - b_4)$ is approximately -0.72, with the spin momentum pointing dominantly along the (1 -1 0) orientation, with the out-of-plane canting of 15°. It is worth noting that this model has one less fitting parameter than noncollinear antiferromagnetism and double spin stripe models. One possible explanation for this anisotropic scattering is that an ordered Jahn-Teller effect may exist on the spinless sublattices [9] that can naturally introduce the tensor with the formula shown in eq. (2) and is compatible with the negative value of $(a_2 - a_4)/(b_2 - b_4)$. Another possibility is that magnetic quadrupole order in the spinless sublattices leads to the anisotropic scattering from the mixture of dipole-quadrupole in eq. (3). If we only consider the spherical symmetry and assume that the local momentum is mainly pointing in the crystallographic *ab* plane, we can define the direction of the local moment as ($x \equiv m_x/|m|$, $y \equiv m_y/|m|$, $z \equiv m_z/|m| = 0$). The scattering tensor for the quadrupole terms can be written as:

$$F_i = \begin{pmatrix} (x^2 - \frac{1}{3})F^2 & (xy)F^2 & 0 \\ (xy)F^2 & (y^2 - \frac{1}{3})F^2 & 0 \\ 0 & 0 & -\frac{1}{3}F^2 \end{pmatrix}, \tag{5}$$

where $F^2$ describes magnetic linear dichroism and is a real value [8]. It is possible that the slight change in the magnetic quadrupole moments further lowers the local symmetry on the spinless sites. For example, consider that $x \to x + \delta, y \to y - \delta$ for $F_2$, and $x \to x - \delta, y \to y + \delta$ for $F_4$, where x≈y, one finds that

$$F_2 - F_4 \approx \begin{pmatrix} 4x\delta F^2 & 0 & 0 \\ 0 & -4y\delta F^2 & 0 \\ 0 & 0 & 0 \end{pmatrix}. \tag{6}$$

By comparing the second term in eq. (3) and eq. (6), one finds that $(a_2 - a_4)/(b_2 - b_4) = \left(-\frac{x}{y}\right) \sim -1$, which is close to the value of -0.72 from our fitting.

## DFT calculation method and results

The spin order and characteristic wave vector of La$_3$Ni$_2$O$_7$ system in ambient Amam phase was further studied by the Density Functional Theory (DFT) [10,11] calculation as implemented in the Vienna ab initio Simulation Package (VASP) [12,13]. The calculations were carried out within the orthorhombic convention cell of the Amam phase as shown in Fig. 1a. The generalized gradient approximation (GGA) of Perdew-Burke-Ernzerhof (PBE) [14] type was adopted for the exchange-correlation functional. The kinetic energy cutoff of plane-wave basis was set to be 500 eV. The Gaussian smearing method with a width of 50 meV was employed for the Fermi surface broadening. The lattice constants and internal atomic positions were fixed at our relaxed results in non-magnetism calculation ($a_0$ = 5.41 Å, $b_0$ = 5.52 Å, $c_0$ = 20.30 Å). An 8 × 8 × 2 $\boldsymbol{k}$-point mesh was used for the Brillouin zone (BZ) sampling of this orthorhombic conventional cell and will be proportionally reduced in other magnetism supercell calculations.

Plentiful collinear spin configurations within the bilayer square lattice were studied. By analysis of our calculation results, we found the spin exchange between two Ni-O layers in this bilayer system consistently favors anti-ferromagnetic. This suggests the stable interlayered anti-ferromagnetic ground state. Therefore, in our subsequent discussions, we will solely concentrate on the spin configurations within a single layer while fixing the interlayered spin configuration at anti-ferromagnetic. Some typical single layered spin configurations in our calculations were exhibited in Fig. S8 a-h and their related energy differences can be found in Table S1. Here the energy reference was set to the energy of non-magnetic state. In Table S1, the Double Stripe2 configuration proposed in Ref. [7] exhibits the lowest energy, indicating the magnetic ground state of $La_3Ni_2O_7$ in the *Amam* phase. The corresponding wave vector is (1/4, 1/4), consistent with our observation on $Q_{SP}$. This result also implies that the PBE functional [14] without other corrections can also capture the partial spin-wave character of this $La_3Ni_2O_7$ system. Nevertheless, we also carried out the Dudarev-type DFT+U calculations [15]. When we set U = 3 eV, we found the magnetism ground state of the system will change to a typical G-AFM state with a characteristic wave vector of (1/2, 1/2). This suggests that the DFT+U strategy might excessively correct the correlation effect on Ni 3*d* electrons, potentially overshadowing other crucial factors. Notably, similar results were also reported in a recent first-principles calculation study on $La_3Ni_2O_7$ for its *Fmmm* phase [16].

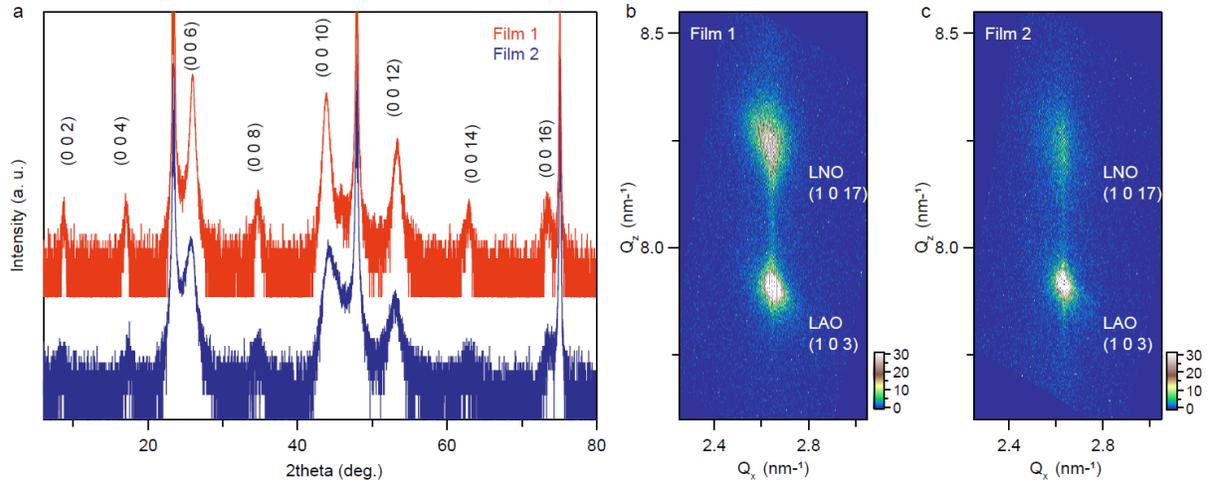

**FIG. S1: Epitaxial growth of high-quality La$_3$Ni$_2$O$_{7-\delta}$ films on LaAlO$_3$ (001) substrates. a**, The out-of-plane X-ray diffraction (XRD) patterns for two La$_3$Ni$_2$O$_{7-\delta}$ films. **b**, **c,** Reciprocal space maps (RSM) of the (1 0 17) reflection for Film 1 and Film 2, respectively.

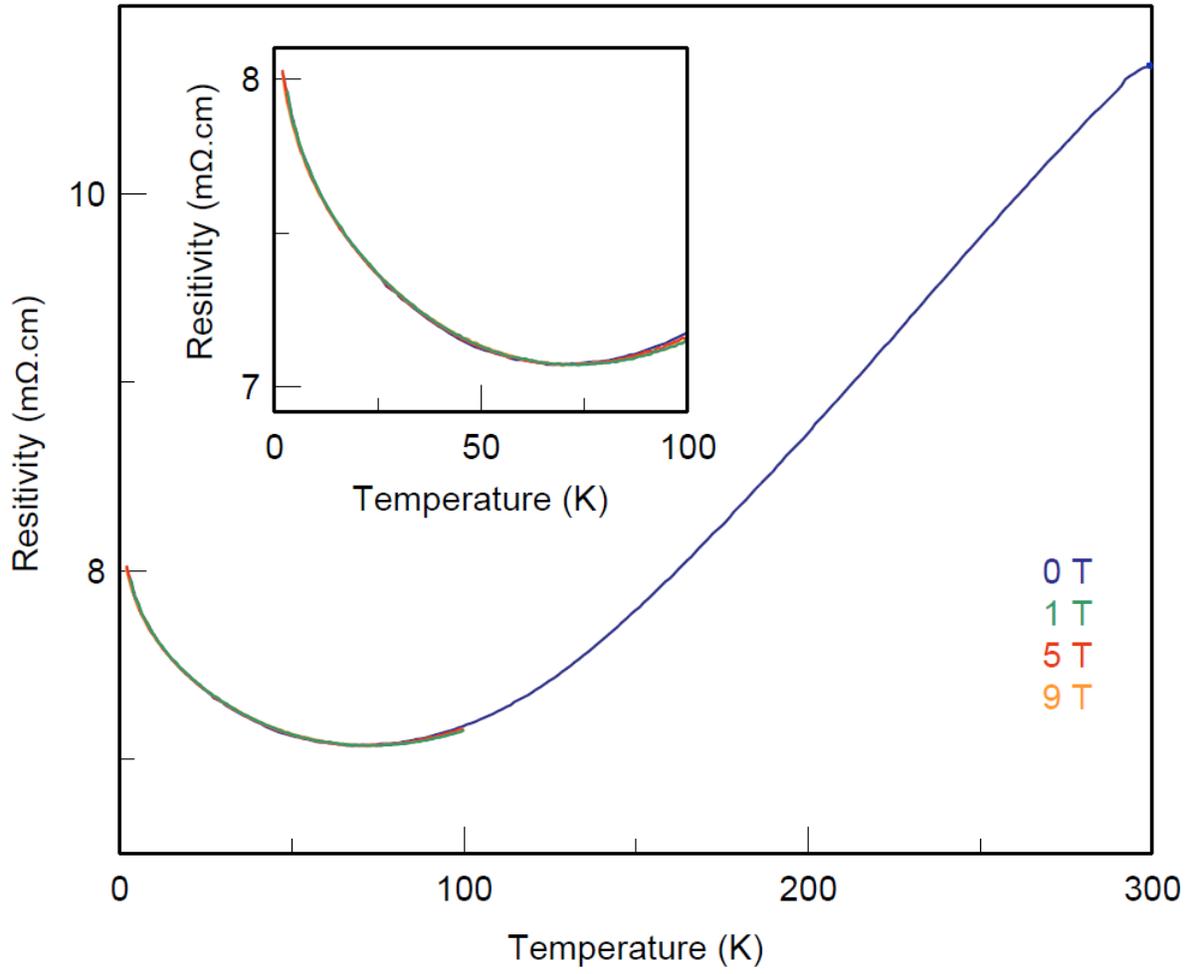

FIG. S2: Resistivity versus temperature under various magnetic fields of a typical $La_3Ni_2O_{7-\delta}$ sample.

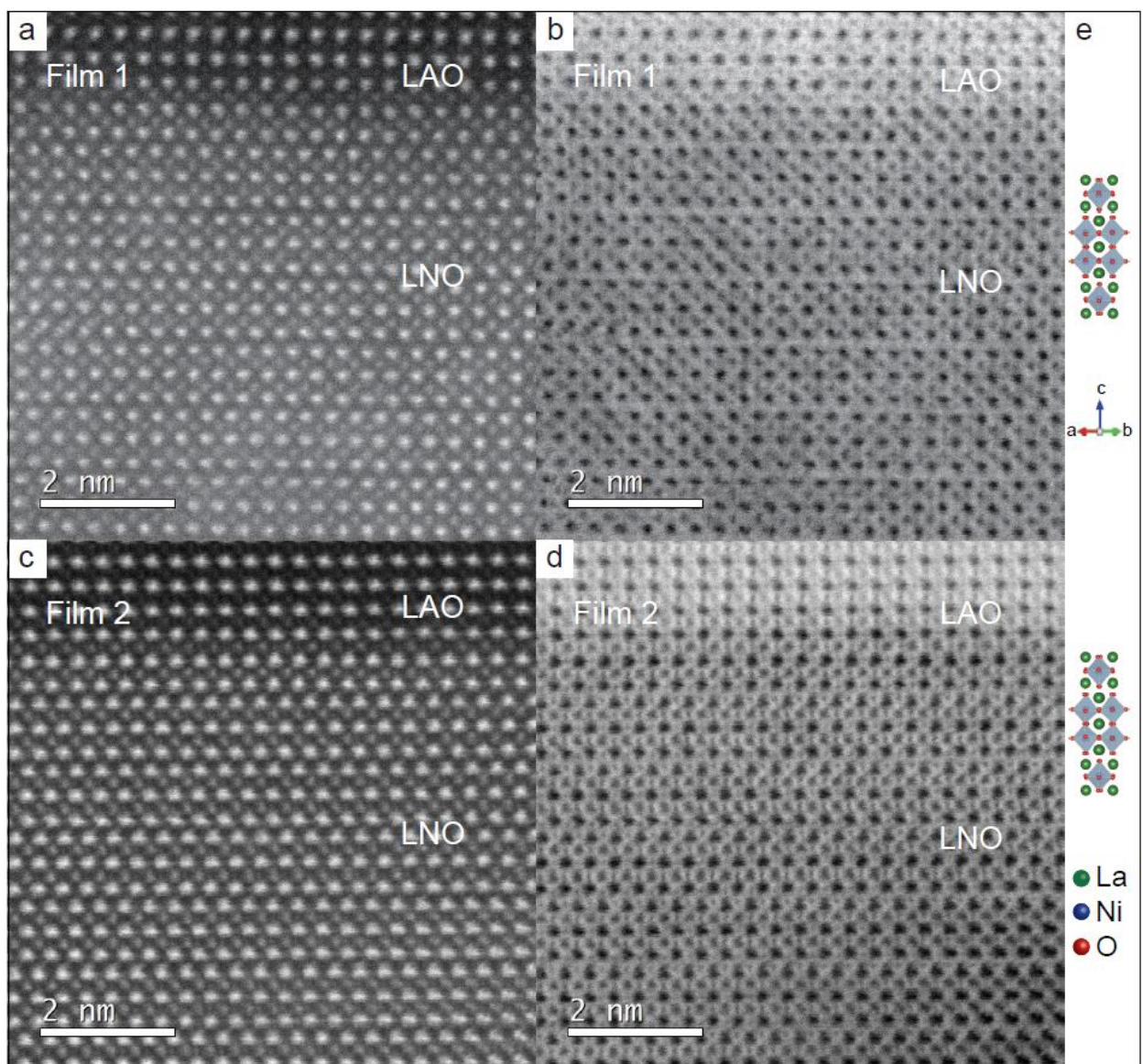

**FIG. S3: Scanning transmission electron microscopy (STEM) images for two La$_3$Ni$_2$O$_{7-\delta}$ films. a (c), b (d),** The high-angle annular dark field (HAADF) STEM and the annular bright field (ABF) STEM images of Flm1 (Film 2), respectively. **e**, The schematics of the crystal structures of the La$_3$Ni$_2$O$_{7-\delta}$ films.

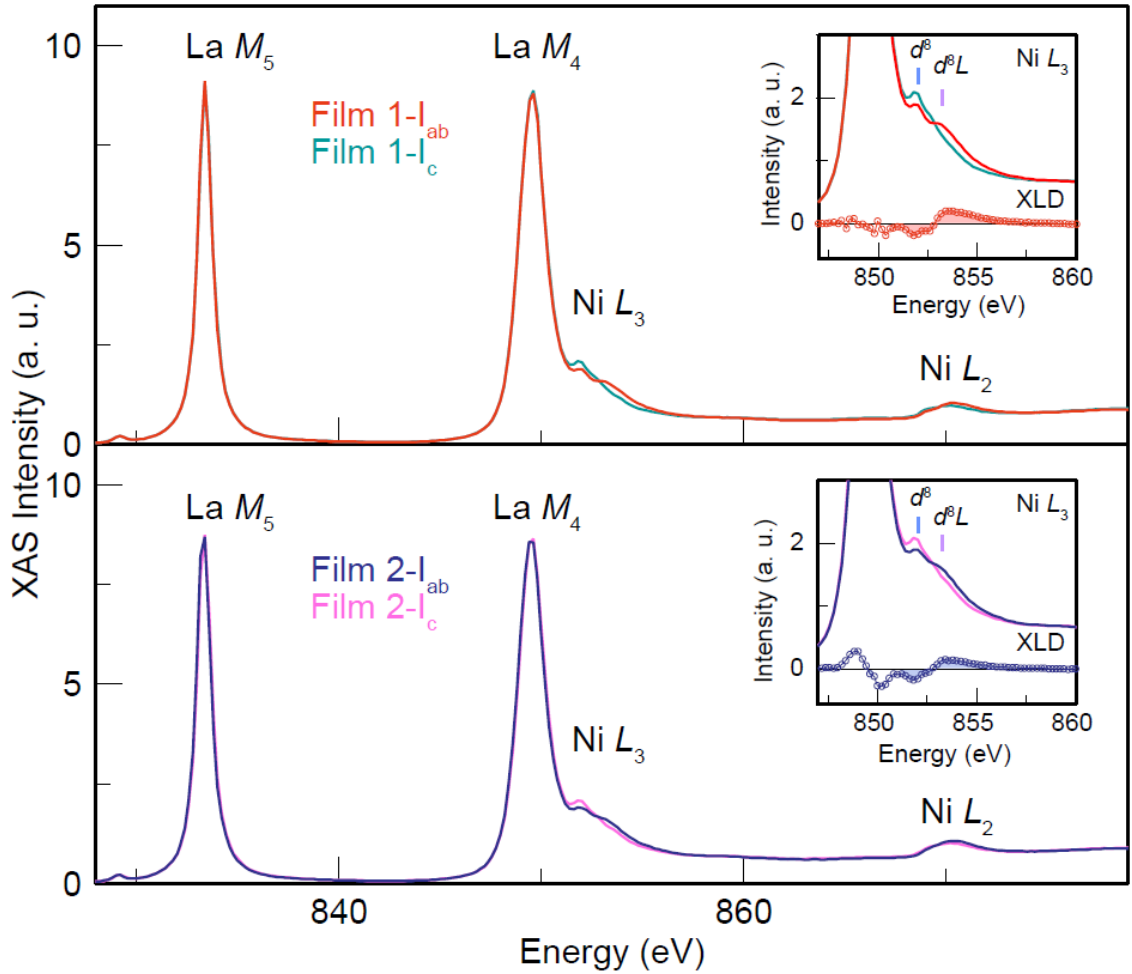

**FIG. S4: XAS of $La_3Ni_2O_{7-\delta}$ films collected at the La-$M$ edge and the Ni-$L$ edge via total electron yield (TEY) mode.** The XAS at the La $M$ edge and the Ni $L$ edge of two $La_3Ni_2O_{7-\delta}$ films with the linear photon polarization perpendicular ($I_{ab}$) and parallel ($I_c$) to the films' $c$ axis. The inset shows an enlarged view of XAS near the Ni-$L_3$ absorption edge along with the XLD analysis.

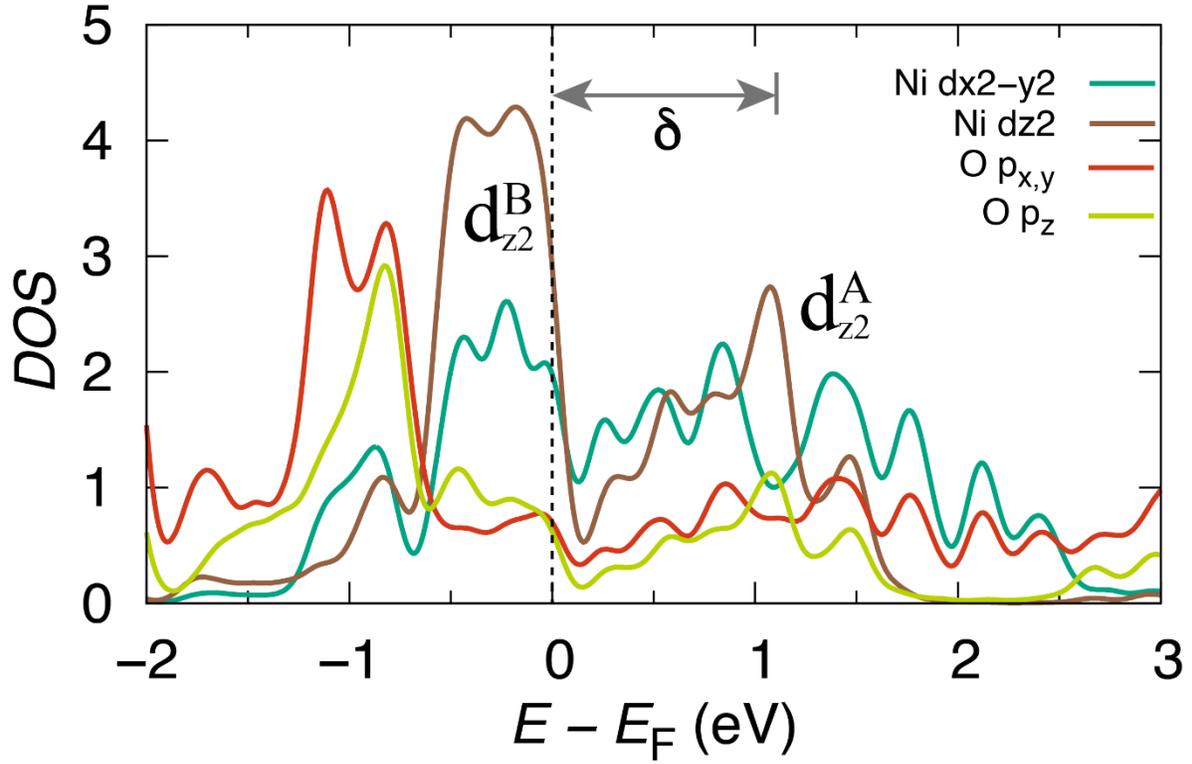

**FIG. S5: Orbital-resolved DOS for bulk La$_3$Ni$_2$O$_7$.** The d$^B_{z2}$ (d$^A_{z2}$) denotes the $d_{z^2}$ interlayer bonding (antibonding) state. The Ni $d_{x^2-y^2}$ orbital shows strong hybridization with O $p_{x,y}$ orbitals, while Ni $d_{z^2}$ exhibits strong hybridization with O $p_z$ orbital.

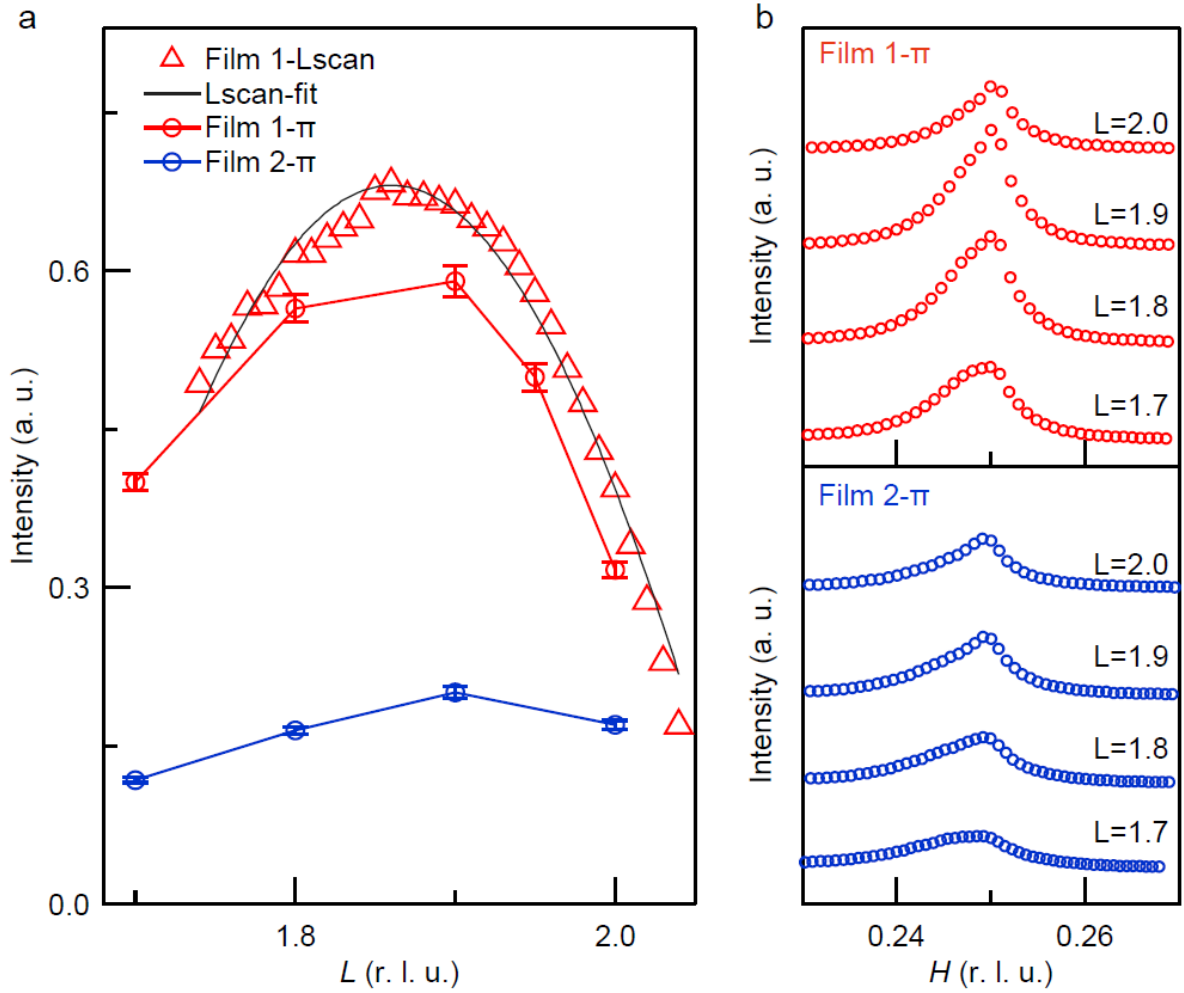

**FIG. S6:** *L* **dependence of the superlattice peak at (1/4, 1/4, *L*) in La$_3$Ni$_2$O$_{7-\delta}$ films. a**, *L* dependence of the scattering intensity measured at 20 K for La$_3$Ni$_2$O$_{7-\delta}$ films with the π-polarized incident photons of energy $E$ = 852.2 eV (near the Ni $L_3$ edge); the fitting to the data using the Gaussian profile is given by the black solid line. **b**, The momentum scans across the wavevector (1/4, 1/4, *L*) with a series of *L* values. The spectra are shifted vertically for clarity.

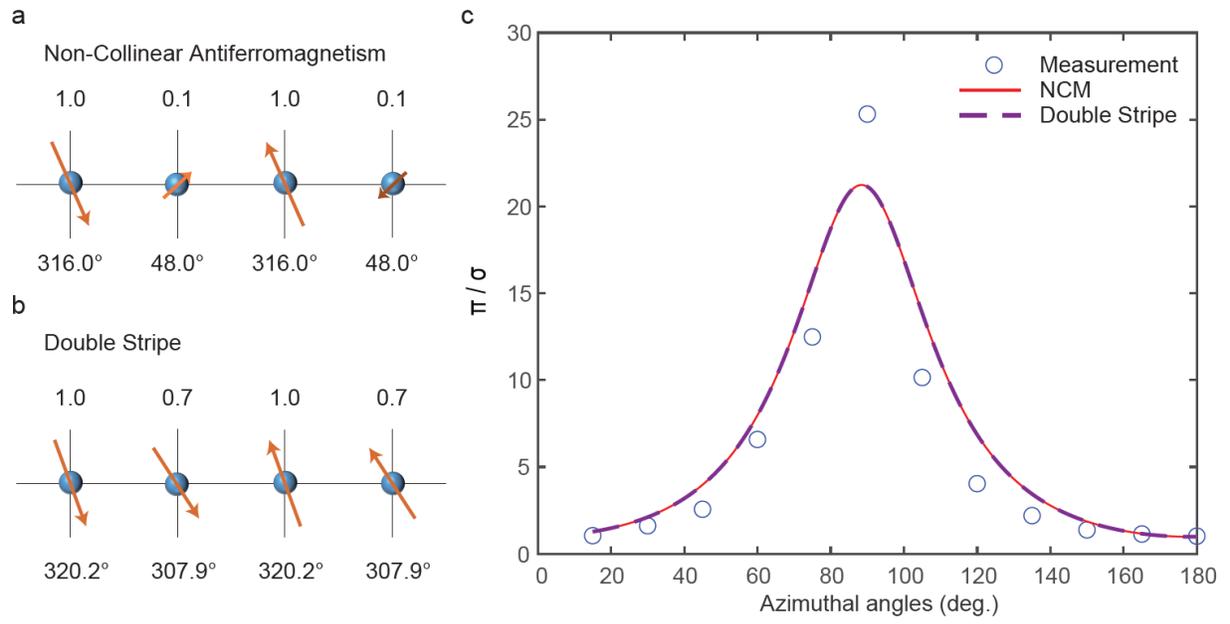

**FIG. S7: Azimuthal angle dependence of the scattered intensity of resonant reflection at (1/4, 1/4, 1.9).** **a**, **b**, Schematics of non-collinear antiferromagnetism (NCM) and double stripe in real space. **c**, The dashed and solid lines represent the best fit to the measurement, corresponding to the non-collinear antiferromagnetism and double stripe spin texture, respectively.

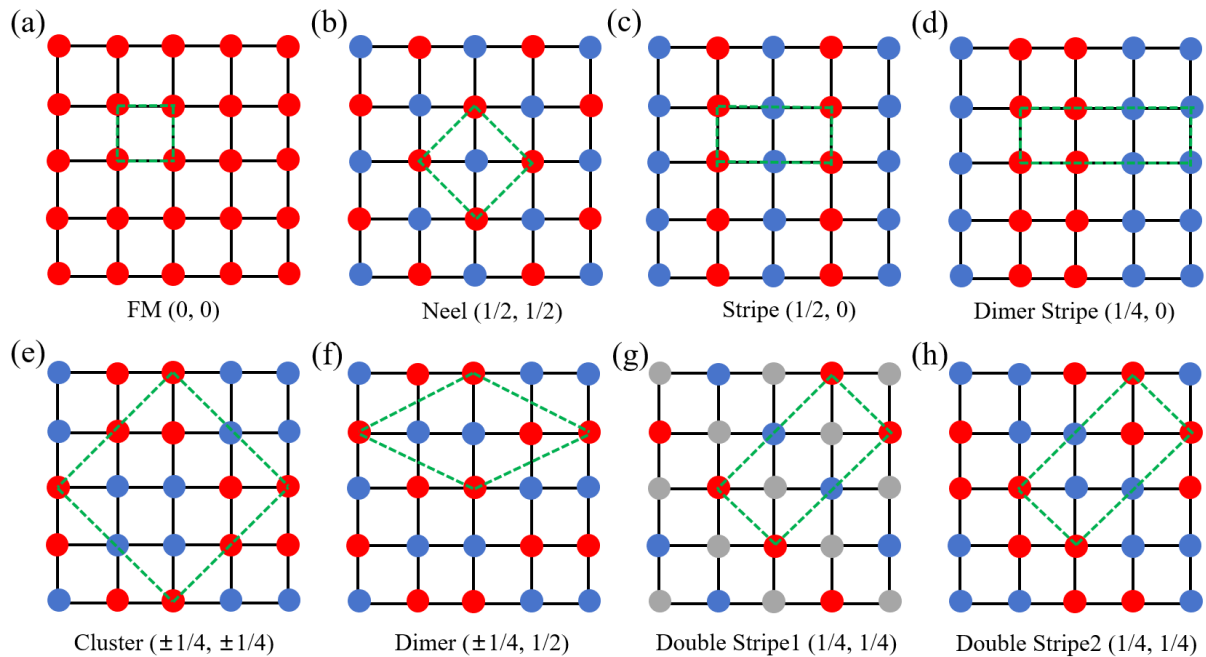

**FIG. S8: Colinear spin configurations within single Ni-O layer. a-h**, The interlayered spin configuration was fixed at anti-ferromagnetic. The red and blue dots signed the Ni atoms with opposite spin directions. The grey dots in panel (g) signed the non-magnetic Ni atoms. The green dotted lines frame out their spin superlattice and the corresponded characteristic wave-vector were also denoted below.

| Magnetism | Energy | Wave vector | Moment |
|---|---|---|---|
| NM | 0.0 | (0, 0) | 0.0 |
| FM | -4.21 | (0, 0) | 0.175 |
| Neel | -20.98 | (1/2, 1/2) | 0.563 |
| Stripe | -32.91 | (1/2, 0) | 0.748 |
| Dimer Stripe | -26.77 | (1/4, 1/2) | 0.748 |
| Dimer | -41.78 | (1/4, 1/2) | 0.713 |
| Cluster | -40.27 | (±1/4, ±1/4) | 0.755 |
| Double Stripe1 | *, -27.69 | (1/4, 1/4) | *, 0.854 |
| Double Stripe2 | -51.87, -42.60 | (1/4, 1/4) | 0.788, 0.754 |

**TABLE S1:** The energy comparison (in units of meV/Ni) between different spin configurations in Fig. S7. The third and fourth column are their characteristic wave-vectors and average magnetic moments (in units of $\mu_B$). The two values in Double Stripe1 and Double Stripe2 configurations respectively correspond to two vertical stripy directions, whose degeneration was broken in the *Amam* phase of $La_3Ni_2O_7$ at ambient pressure. The asterisk in Double Stripe1 denotes the calculation doesn't converge to our specified spin configuration.